# Crystallographic Disorder and Strong Magnetic Anisotropy in $Dy_3Pt_2Sb_{4.48}$

*Terry Paske,[a] Yingdong Guan,[b] Chaoguo Wang,[a] Curtis Moore,[c] Zhiqiang Mao,[b] Xin Gui [a*]*

[a] Department of Chemistry, University of Pittsburgh, Pittsburgh, PA, 15260, USA
[b] Department of Physics, Pennsylvania State University, University Park, PA, 16801, USA
[c] Department of Chemistry and Biochemistry, The Ohio State University, Columbus, OH, 43210, USA

## *Abstract*

We report the crystal growth and characterization of a rare-earth-containing material, $Dy_{3.00(1)}Pt_2Sb_{4.48(2)}$. This compound possesses a similar structure to the previously reported $Y_3Pt_4Ge_6$, but lacks two layers of Pt atoms. Crystallographic disorder was found in $Dy_{3.00(1)}Pt_2Sb_{4.48(2)}$. Additionally, the Dy-Dy framework was found to have both square net and triangular lattices. $Dy_{3.00(1)}Pt_2Sb_{4.48(2)8}$ was determined to be antiferromagnetically ordered around ~15 K while a competing antiferromagnetic sublattice also exists at lower temperature. Strong magnetic anisotropy was observed and several metamagnetic transitions were seen in the hysteresis loops. Furthermore, the Curie-Weiss fitting revealed unusually small effective moment of Dy, which is far below the expected value of $Dy^{3+}$ (10.65 $\mu_B$). This material might provide a new platform to study the relation between crystallographic disorder and magnetism.

## Introduction

Rare-earth-based materials are of great importance for the investigation of frustrated magnets since they possess highly localized moments and their ability to crystallize into various structural types. It is well known that magnetic frustration commonly results from the competing antiferromagnetic (AFM) interactions between the nearest neighbors or even the next nearest neighbors.[1,2] To allow such competition, geometric frustration of the sublattice of magnetic atoms is one of the most important routes to realizing frustrated magnetism. The most widely investigated frustrated structural motifs include triangular lattice (YbMgGaO$_4$),[3–6] Kagome lattice (ZnCu$_3$(OH)$_6$Cl$_2$),[3,7–9] and pyrochlore/breathing pyrochlore lattice (Ce$_2$Sn$_2$O$_7$/Ba$_3$Yb$_2$Zn$_5$O$_{11}$).[10–14] These structural motifs are all composed of triangles. Ideally, equilateral triangles should exist in such systems and their frustration can be "isotropic". However, square lattices were also found to be great motifs for magnetic frustration since it can be seen as two edge-shared isosceles right triangles.[15–17] Thus when the competing exchange coupling between the nearest neighbors ($J_1$) and the next nearest neighbors ($J_2$) fall into an appropriate region, magnetic frustration can be expected.[18]

Y$_3$Pt$_4$Ge$_6$ was reported to crystallize in the $P\,2_1/m$ space group.[19] In recent years, the crystal structure and magnetic properties of polycrystalline Ln$_3$Pt$_4$Ge$_6$ (Ln = Y, Ce, Pr, Nd, Sm, Gd-Dy) were reinvestigated[20,21] and isostructural Yb$_3$Pt$_4$Si$_{6-x}$[22] was also discovered. What is interesting about Ln$_3$Pt$_4$Ge$_6$ is that the atoms on the Ln site construct frameworks of both tilted square lattice and triangular lattice. The magnetic properties measurements of these materials revealed that only Gd$_3$Pt$_4$Ge$_6$ and Tb$_3$Pt$_4$Ge$_6$ were antiferromagnetically ordered while others were paramagnetic.[20] Pure phase of polycrystalline Dy$_3$Pt$_4$Ge$_6$ could not be synthesized and thus its magnetic properties were not reported.[20]

Here, we replaced Ge with Sb and obtained mm-sized single crystals of Dy$_{3.00(1)}$Pt$_2$Sb$_{4.48(2)}$ with crystallographic disorder on both Dy and Sb sites. Dy$_{3.00(1)}$Pt$_2$Sb$_{4.48(2)}$ appears to be a new material which crystallizes in a similar structure type as RE$_3$Pt$_4$Ge$_6$. By performing magnetic properties and electrical transport measurements, we observed complex magnetic behavior in Dy$_{3.00(1)}$Pt$_2$Sb$_{4.48(2)}$ including strong magnetic anisotropy, competing AFM sublattices and multiple metamagnetic transitions. We believe that Dy$_{3.00(1)}$Pt$_2$Sb$_{4.48(2)}$ will provide a platform for investigating the relation between crystallographic disorder and magnetism as well as frustrated magnetism.

*Experimental Details*

**Single Crystal Growth:** The self-flux method with excess Sb was employed to obtain single crystals of $Dy_{3.00(1)}Pt_2Sb_{4.48(2)}$. Dy powder (99.9%, ~400 mesh, Alfa Aesar), Pt powder (>99.9%, ~325 mesh, Thermo Scientific) and Sb powder (99.5%, ~100 mesh, Thermo Scientific) in the molar ratio of 3:2:50 were mixed in an alumina crucible in an Ar-filled glovebox. This crucible was put into an quartz tube, with quartz wool and some glass pieces above to assist the separation of the Sb flux from the crystals after cooling. The tube was purged with argon and then evacuated (<5×10$^{-3}$ Torr) and sealed. It was heated in a box furnace to 900 °C and held for 12 hours followed directly by another heat treatment at 1100 °C for two days. After one day of heating at 1100 °C, the tube was gently agitated, during which, the temperature within did not drop below 1000 °C. The furnace was slowly cooled at the rate of 1 °C per hour to 750 °C, where it remained for a few hours then the Sb flux was centrifuged out. Water quenching the tube afterwards solidified the Sb flux in the wool plug for simple separation from the crystals. Striped crystals with dimensions up to 0.1 x 0.25 x 3 mm$^3$ were obtained, as shown in Figure 1(a). The crystals are stable in air.

**Structure and Phase Determination:** The single crystal X-ray diffraction studies were carried out on a Bruker Kappa Photon III CPAD diffractometer equipped with Mo K$_\alpha$ radiation ($\lambda$ = 0.71073 Å). A 0.125 x 0.093 x 0.043 mm piece of a metallic silver block was mounted on a MiTeGen Micromount with Paratone 24EX oil. Data were collected in a nitrogen gas stream at 125(2) K using $\phi$ and $\varpi$ scans. Crystal-to-detector distance was 50 mm using variable exposure time (2-5 s) depending on θ with a scan width of 1.0°. Bruker SMART software with corrections for Lorentz and polarization effects included was utilized for data requisition. Numerical absorption correction was applied based on crystal-face-indexing using *XPREP*. The direct method and full-matrix least-squares on F$^2$ procedure within the SHELXTL package were employed to solve the crystal structure.[23,24] Powder X-ray diffraction patterns on crushed crystals, obtained with a Bruker D2 Phaser diffractometer with Cu K$\alpha$ radiation and a LYNXEYE-XE detector, were consistent with the structure determined by single crystal diffraction.

**Scanning Electron Microscopy (SEM)-Energy-Dispersive X-ray Spectroscopy (EDS):** The high vacuum SEM (ThermoFisher Apreo HiVac with EDAX Elite 150 SDD EDS detector) was used to determine the chemical composition. Samples were mounted on carbon tape before loading into the SEM chamber. Multiple points and areas were examined from each sample to get the Dy: Pt: Sb ratio. Samples were analyzed at 20 kV, and the spectra were collected for 100 seconds to get the chemical composition via APEX EDS software.

**Physical Property Measurement:** Oriented single crystal samples were utilized for physical properties measurement. The magnetic properties were measured on a Quantum Design SQUID magnetometer from 2-300 K under the external magnetic field of 0-7 T. The electrical transport measurement was conducted on a Quantum Design PPMS from 2.5 K to 300 K under the external magnetic field of 0-9 T using a four-probe method.

**Electronic Structure Calculations:** Calculations of Crystal Orbital Hamiltonian Population (COHP) was performed by Tight-Binding, Linear Muffin-Tin Orbital-Atomic Spheres Approximation (TB-LMTO-ASA) using the Stuttgart code.[25–27] A mesh of 1000 $k$ points was used to generate all integrated values.[28] The number of irreducible k-points is 70. Energy difference of 0.05 meV was set as the convergence criterion. In the ASA method, space was filled with overlapping Wigner-Seitz (WS) spheres. The symmetry of the potential was treated as spherical in each WS sphere with a combined correction on the overlapping part. The WS radii were: 3.67 Å for Dy; 2.59 Å for Pt; and 3.30 Å for Sb. Empty spheres were required for the calculation, and the overlap of WS spheres was limited to no more than 16%.

Band structure and density of states (DOS) were calculated by using WIEN2k, which employs the full-potential linearized augmented plane wave method (FP-LAPW) with local orbitals implemented.[29,30] The electron exchange-correlation potential used was the generalized gradient approximation.[31] The conjugate gradient algorithm was applied. Reciprocal space integrations were completed over a 5×10×3 Monkhorst-Pack $k$-point mesh.[32] Orbital potentials (U = 4 eV) were employed for Dy $f$ electrons. With these settings, the calculated total energy converged to less than 0.1 meV per atom.

## Results and Discussion

**Crystal Structure Determination:** By performing single-crystal X-ray diffraction, the formula of the striped crystals was determined to be $Dy_{3.00(1)}Pt_2Sb_{4.48(2)}$, which crystallizes in the monoclinic space group $P\,2_1/m$ (No. 11, $mP19$) and is found to be a new material in the Dy-Pt-Sb ternary system. The crystallographic data, including atomic positions, site occupancies, and refined anisotropic displacement parameters (and equivalent isotropic thermal displacement parameters) are listed in Tables 1, 2 and 3. In our crystallographic model, the best solution emerged when multiple disordered sites were introduced, as shown in Figure 1(b) where 12 disordered sites can be found and six of them (Dy1, Sb1, Sb1' and Dy2, Sb2, Sb2') are symmetrically equivalent with the rest correlated by the $2_1$ symmetry. Here we correlate Dy1 with Sb1 and Sb1' as well as Dy2 with Sb2 and Sb2', indicating that Dy1(2) site exists only when Sb1(1) and Sb1'(2') appear. The interatomic distances between Dy1/Dy2 and Sb2 (Sb2')/Sb1 (Sb1') are ~1.29 Å. These are not possible bond lengths as they are far below 50% of the common Dy-Sb bond lengths, for example, 3.0660 (2) Å in DySb[33,34]. Moreover, by removing either set of Dy and Sb atoms, a large residual peak and hole can appear, as can be seen in Table S1. Therefore, in order to obtain reasonable crystallographic refinement results, two constraints must be applied to the crystallographic model: 1. The sum of the occupancies of Dy1 and Dy2 was set to be 100%; 2. The occupancies of sites Sb1, Sb1', Sb2 and Sb2' were set to be correlated and relaxed. In such condition, Sb1 and Sb1' turned to possess higher occupancies (~20% for each site) than Sb2 and Sb2' (~4% for each site) and the interatomic distances between Dy1-Sb1 (3.023 (3) Å), and Dy2-Sb2 (3.02 (2) Å) pairs are comparable with literature about compounds in Dy-Sb binary system and elemental Sb.[33-34] The bond length of Sb1-Sb1' (2.579 (4) Å) and Sb2-Sb2' (2.60 (3) Å) are smaller than literature[35]. Considering that such Sb pairs exist as split interstitial, which can usually lead to short bond length, the short Sb-Sb bond length in this work is reasonable. Under such constraints, the crystallographic disorder with very short Dy-Sb distances can thus be interpreted in a reasonable way that the Dy1, Sb1 and Sb1' sites are occupied alternately with the Dy2, Sb2 and Sb2' sites based on the site occupancies shown in Table 2. A comparison of crystallographic refinement results with different models has also been summarized in Table S1. The inclusion of Sb2(Sb2') and Dy2 significantly improves the results. As shown in Figure 1(c), sites Dy1 and the Sb1 and Sb1' pair make up setting #1, and Dy2 and the Sb2 and Sb2' pair make up setting #2. Same coordination environment of Dy can be found in both settings. Two types of Dy@Sb$_8$ polyhedra are shown on the left of Figure 1(d). Additionally, when coordinated with Pt atoms, the partially-occupied Dy forms Dy@Pt$_6$ trigonal prisms while fully-occupied Dy constructs Dy@Pt$_2$ zig-zag chain. When bonded together, the fully-occupied Dy atoms exhibits a distorted square net framework, as shown in Figure 1(e). Moreover, partially-occupied Dy1 atoms construct a puckered triangular lattice. The Dy framework can be seen as extending two-dimensionally within the *ab* plane and stacking along the *c* axis.

To confirm the chemical composition, the SEM-EDS measurements were performed on two different crystals from different batches. The EDS results averaged from five points are shown in Table S2 in the Supplementary Information. The average formula was determined to be Dy$_{3.1(3)}$Pt$_{2.0(7)}$Sb$_{3.8(4)}$ when normalized to the concentration of Pt, which is consistent with the results from the single-crystal X-ray diffraction measurement. Moreover, powder X-ray diffraction pattern of the crystals is illustrated in Figure S1, which implies high purity of the crystals. A preferred orientation of (00l) is clearly seen in the powder XRD pattern.

**Structural Relations and Bonding Analysis:** When considering several repeats of the unit cell along the *c* axis, certain structural motifs within previously reported structures can be identified. Both YIrGe$_2$ and YbSb$_2$ have a top and bottom motif that are symmetrical about a mirror plane through the center of the unit cell horizontally, Figure 2(a) (Left) and (Right) respectively. The setting#1 of Dy$_{3.00(1)}$Pt$_2$Sb$_{4.48(2)}$ can be seen as the intergrowth of alternating slabs of YIrGe$_2$ and YbSb$_2$ along the *c* axis, shown in Figure 2(a) (Middle). The adjacent YIrGe$_2$ and YbSb$_2$ slabs share a layer of Dy and Sb in Dy$_{3.00(1)}$Pt$_2$Sb$_{4.48(2)}$. As for the setting#2 of Dy$_{3.00(1)}$Pt$_2$Sb$_{4.48(2)}$, the order of the top and bottom halves of YIrGe$_2$ switch, which is not shown here. This repeating motif in Dy$_{3.00(1)}$Pt$_2$Sb$_{4.48(2)}$ is similar to repeat slabs in Y$_3$Pt$_4$Ge$_6$ for which previous reports[19] claimed to be built from alternating YIrGe$_2$-type and ThCr$_2$Si$_2$-type[36,37] slabs. However, as can be seen in Figure 2(b), the alternating layer in Y$_3$Pt$_4$Ge$_6$ contains two layers of Pt atoms which are missing in Dy$_{3.00(1)}$Pt$_2$Sb$_{4.48(2)}$.

To better understand how the crystallographic disordering affects the structure of Dy$_{3.00(1)}$Pt$_2$Sb$_{4.48(2)}$, COHP calculations were performed for both settings. The relevant chemical bonds for disordered Dy1 and Dy2 are visualized in Figure 3(b) & 3(c) respectively. All bond lengths were set to be no longer than 3.5 Å. Figure 3(a) presents the COHP curves for both settings where the positive part of the x axis indicates bonding interaction, and the negative part shows anti-bonding interaction. Similar behaviors can be found for Dy1(Dy2)-Sb1(Sb2) and Dy1(Dy2)-Sb1'(Sb2') as well as Dy1(Dy2)-Pt1 and Dy1(Dy2)-Pt2. The disordered Dy in both settings shows nearly all anti-bonding features when bonded with the surrounding atoms within the visible energy range (-4 eV to 2 eV). Such behavior indicates that the disordered Dy1 and Dy2 sites both destabilize the structure from -4 eV to 2 eV, which might explain why they are both partially occupied. More COHP curves for Dy-Pt, Dy-Sb, Pt-Sb and Sb-Sb bonds can be found Figure S2 in the Supporting Information, which demonstrates significant anti-bonding interaction above ~ -2.5 eV.

**Magnetic Characterization:** Magnetic property measurements were performed on $Dy_{3.00(1)}Pt_2Sb_{4.48(2)}$ single crystals with the magnetic field applied along two directions, perpendicular to the *ab* plane (H⊥*ab*), and parallel to the *a* axis (H//*a*). The orientation of a representative stripe-like crystal is shown in the inset of Figure 4(a). The temperature-dependence of magnetic susceptibility ($\chi$) and inverse $\chi$ under an applied magnetic field of 1000 Oe from 2 to 300 K are shown in Figure 4(a) and 4(c). Both curves were collected under field-cooling (FC) history. We observed distinct magnetic anisotropy from the data presented in Figure 4(a-b). When the field (H) is perpendicular to the *ab* plane (Figure 4(a)), an AFM transition can be found around $T_{N1}$ ~ 17 K while another peak is seen at $T_{N2}$ ~ 4.5 K. The magenta curve in Figure 4(a) shows the temperature dependence of $1/\chi$. When the external magnetic field is parallel to the *a* axis, the $\chi$ vs T curve exhibits different behavior. As shown in the main panel of Figure 4(c), the magnetic susceptibility increases with decreasing temperature until a kink is seen at ~ 15.9 K, indicating the spin easy axis is within the ab-plane. Upon further cooling, another sharp peak appears at ~3.9 K. We fitted the $1/\chi$ data using the modified Curie-Weiss (CW) law,

$$\chi = \chi_0 + \frac{C}{T-\theta_{CW}},$$

where $\chi$ is the magnetic susceptibility of the material, $\chi_0$ and C are independent of temperature (the latter related to the effective moment and the former to the core diamagnetism and temperature independent paramagnetic contributions such as Pauli paramagnetism), and $\theta_{CW}$ is the Curie-Weiss constant. The best fit for both directions was obtained in the temperature range of 150-300 K. $\theta_{CW}$ obtained from the fit is ~ -11.1 (2) K when H⊥*ab* and ~ -17.9 (1) K when H//*a*; the negative sign suggests antiferromagnetic interaction within the fitted temperature range. The effective moment ($\mu_{eff}$) per Dy is calculated by $\mu_{eff}/Dy = \sqrt{8C}/3$ $\mu_B$, which leads to 6.98 (1) $\mu_B$/Dy when H⊥*ab* and 5.74 (1) $\mu_B$/Dy when H//*a*, both of which are far below the expected value of $Dy^{2+}$ (10.6 $\mu_B$), $Dy^{3+}$ (10.65 $\mu_B$) and $Dy^{4+}$ (9.72 $\mu_B$). The small $\mu_{eff}$ may be due to the intermediate *f*-electron configuration of Dy in our material.[38] The difference between $\mu_{eff}$ from two directions may originate from the strong magnetic anisotropy.

To further understand the magnetic behaviors of $Dy_{3.00(1)}Pt_2Sb_{4.48(2)}$, the hysteresis loops were measured as shown in Figure 4(b) and 4(d). For H⊥*ab*, several metamagnetic transitions can be seen in Figure 4(b). To better visualize it, a plot from 0 to 7 T for the first cycle of the loop and its first derivative is placed on the bottom-right corner of Figure 4(b). Four obvious peaks at $H_{M1} \sim 0.25$ T, $H_{M2} \sim 1.1$ T, $H_{M3} \sim 3.5$ T and $H_{M4} \sim 5.1$ T can be found in the derivative curve under 2K, which correspond to four metamagnetic transitions. The other inset on the top-left corner is the enlarged region between -1.5 T and 1.5 T for the loop. No remanent magnetization can be found, and it indicates that there is no significant ferromagnetic component in $Dy_{3.00(1)}Pt_2Sb_{4.48(2)}$, i.e., spins are not canted. Under 2 K, the magnetization at 7 T is ~7.7 $\mu_B$/Dy. When the temperature is raised up to 30 K, a nearly linear relation can be observed. Similarly, the measurement of the hysteresis loop on the other field direction (H//*a*) was also performed. Not surprisingly, strong magnetic anisotropy leads to completely different features for H//*a*. No obvious metamagnetic transition can be seen in Figure 4(d), instead, two slop changes for the hysteresis loop measured under 2 K are observed, evidenced by the magenta dM/dH curve in the inset. The hysteresis loop does not show any trend to saturate when H//*a* at 2 K.

**Electrical Transport Measurements:** Figures 5(a) and 5(b) illustrate the temperature-dependent resistivity and magnetoresistance of $Dy_{3.00(1)}Pt_2Sb_{4.48(2)}$. When no magnetic field is applied, the electrical resistivity of the material decreases with decreasing temperature, indicating the metallic nature, which is further supported by the electronic structure shown in Figure S2. A kink is found ~ 15.5 K in the main panel of Figure 5(a), which reflects the magnetic transition observed at $T_{N1}$ in Figure 4(a) and 4(c). When plotting the first derivative of the ρ vs T curve, two features at ~5 K ad ~9 K can be found. The 5-K transition corresponds with the magnetic transition at $T_{N2}$ while the 9-K transition might originate from another magnetic transition that was not clearly observed in our magnetic measurement. When the magnetic field is applied perpendicularly to the *ab* plane, two peaks, one at ~3.5 T and one at ~5.1 T, can be observed for the magnetoresistance curve at 2.5 K in Figure 5(b). The two peaks are consistent with the two metamagnetic transitions seen in Figure 4(b) at $H_{M3}$ and $H_{M4}$. Although more features are observed under lower field in the magnetic hysteresis loop for H⊥*ab*, no obvious transitions can be found in the magnetoresistance measurements. This is primarily due to the intrinsically low resistivity of the sample and, thus, the low-field metamagnetic transitions are possibly obscured by the noise. Interestingly, when the system is warmed up to 6 K, which is above the $T_{N1}$, the peaks at $H_{M3}$ and $H_{M4}$ disappear while a broad peak emerges at ~4.5 T which persists until 15 K and disappears at 50 K. The broad peak may originate from another metamagnetic transition corresponding with $T_{N1}$.

Based on the findings above, a brief discussion about the magnetic properties of $Dy_{3.00(1)}Pt_2Sb_{4.48(2)}$ shall be made here. Two magnetic transitions are found in the temperature-dependent magnetic susceptibility curves for both directions at $T_{N1}$ ~3.8 K and $T_{N2}$ ~15 K. However, the higher-temperature one is more predominant when H⊥*ab* while when H//*a*, the lower-temperature one prevails. Considering the distinct fitted $θ_{CW}$, we speculate that there are two AFM sublattices in $Dy_{3.00(1)}Pt_2Sb_{4.48(2)}$ which correspond to $T_{N1}$ and $T_{N2}$ in χ vs T curves, i.e., AFM1 that corresponds with $T_{N1}$ and AFM2 that corresponds with $T_{N2}$. Complex interactions between the two AFM sublattices may lead to the multiple metamagnetic transitions in the hysteresis loops at 2 K. Supported by the transport measurement, the two major metamagnetic transitions at $H_{M3}$ and $H_{M4}$ are correlated with the AFM2 sublattice. Besides the speculation above, another possibility of the magnetic structure could be that one Dy sublattice is responsible for different interaction and coupling constants along different crystallographic directions. In the meantime, the other Dy sublattice results in different magnetic orderings with decreasing temperature, such as spin-reorientations etc. Based on the complex observation further studies including X-Ray Magnetic Circular Dichroism (XMCD) measurements need to be conducted to better interpret the magnetic structure of $Dy_{3.00(1)}Pt_2Sb_{4.48(2)}$, especially the relation between the magnetic structure and the two settings of disordered Dy and Sb.

## *Conclusions*

In this paper, motivated by the previously reported material type, $Y_3Pt_4Ge_6$, with both square net and triangular lattices of rare-earth elements, we synthesized mm-sized single crystals of a new material $Dy_{3.00(1)}Pt_2Sb_{4.48(2)}$, which crystallizes in a $Y_3Pt_4Ge_6$-type-related structure. Based on the magnetic properties' measurements, we observed two transitions for AFM ordering at ~15 K and ~4 K, which might originate from two competing AFM sublattices. Furthermore, strong magnetic anisotropy and metamagnetic transitions were found for $Dy_{3.00(1)}Pt_2Sb_{4.48(2)}$. Therefore, further studies including XMCD measurements should be carried out to better understand the complex spin structure of this material. An unusual valence state of Dy was also seen extrapolated from the Curie-Weiss fitting for the effective moment, which deserves follow-up investigations. The magnetic complexity and crystallographic disorder of $Dy_{3.00(1)}Pt_2Sb_{4.48(2)}$ along with the availability of single crystals provides a platform for investigating magnetism in such structural motifs when Dy is substituted by other rare-earth elements.

## *Associated Content*

Supporting Information

The Supporting Information is available free of charge at XXX.

Crystallographic data of the title compound under different constraints. The elemental analysis based on EDS. The powder X-ray diffraction pattern. The proposed magnetic phase diagram.

## *Author Information*

Corresponding Author: xig75@pitt.edu

Notes: The authors declare no competing financial interest.

## *Acknowledgements*

T.L.P., C.W. and X.G. are supported by the startup fund from the University of Pittsburgh and the Pitt Momentum Fund. Y.D.G. and Z.Q.M. acknowledge the support by the US Department of Energy under grants DE-SC0019068 and DE-SC0014208 (support for magnetic and transport measurements).

**Table 1.** Single crystal structure refinement for $Dy_{3.00(1)}Pt_2Sb_{4.48(2)}$ at 273 (2) K.

| Refined Formula | $Dy_{3.00(1)}Pt_2Sb_{4.48(2)}$ |
|---|---|
| F.W. (g/mol) | 1423.14 |
| Space group; Z | $P2_1/m$; 2 |
| $a$ (Å) | 8.6252 (8) |
| $b$ (Å) | 4.3109 (3) |
| $c$ (Å) | 12.968 (1) |
| $\beta$ (°) | 99.609 (3) |
| V (Å$^3$) | 475.43 (7) |
| Extinction Coefficient | 0.00037 (5) |
| θ range (°) | 3.090-31.902 |
| No. reflections; $R_{int}$ | 23877; 0.0329 |
| No. independent reflections | 1699 |
| No. parameters | 75 |
| $R_1$: $\omega R_2$ ($I>2\delta(I)$) | 0.0272; 0.0432 |
| Goodness of fit | 1.151 |
| Diffraction peak and hole (e$^-$/ Å$^3$) | 2.954; -2.613 |

**Table 2.** Atomic coordinates and equivalent isotropic displacement parameters for $Dy_{3.00(1)}Pt_2Sb_{4.48(2)}$ at 273 (2) K. ($U_{eq}$ is defined as one-third of the trace of the orthogonalized $U_{ij}$ tensor (Å$^2$))

| Atom | Wyck. | Occ. | x | y | z | $U_{eq}$ |
|---|---|---|---|---|---|---|
| Pt1 | 2e | 1 | 0.52636 (4) | ¼ | 0.61894 (3) | 0.0044 (1) |
| Pt2 | 2e | 1 | 0.03314 (4) | ¼ | 0.61904 (3) | 0.0045 (1) |
| Dy1 | 2e | 0.830 (1) | 0.23486 (7) | ¼ | 0.43955 (5) | 0.0042 (1) |
| Sb1 | 2e | 0.200 (3) | 0.4155 (4) | ¾ | 0.5623 (3) | 0.0051 (7) |
| Sb1' | 2e | 0.201 (3) | 0.1156 (4) | ¾ | 0.5621 (3) | 0.0051 (7) |
| Dy2 | 2e | 0.170 (1) | 0.2648 (3) | ¾ | 0.5606 (2) | 0.0042 (1) |
| Sb2 | 2e | 0.039 (2) | 0.384 (2) | ¼ | 0.439 (1) | 0.0051 (7) |
| Sb2' | 2e | 0.039 (2) | 0.083 (2) | ¼ | 0.437 (1) | 0.0051 (7) |
| Dy3 | 2e | 1 | 0.57655 (6) | ¾ | 0.81070 (4) | 0.0050 (1) |
| Dy4 | 2e | 1 | 0.07890 (6) | ¾ | 0.81115 (4) | 0.0050 (1) |
| Sb3 | 2e | 1 | 0.12475 (7) | ¼ | 0.99840 (5) | 0.0053 (1) |
| Sb4 | 2e | 1 | 0.37569 (7) | ¾ | 0.00186 (5) | 0.0054 (1) |
| Sb5 | 2e | 1 | 0.31256 (7) | ¼ | 0.74994 (5) | 0.0051 (1) |
| Sb6 | 2e | 1 | 0.81150 (7) | ¼ | 0.74621 (5) | 0.0049 (1) |

**Table 3.** Anisotropic thermal displacement parameters for $Dy_{3.00(1)}Pt_2Sb_{4.48(2)}$.

| Atom | $U_{11}$ | $U_{22}$ | $U_{33}$ | $U_{23}$* | $U_{13}$* | $U_{12}$* |
|---|---|---|---|---|---|---|
| Pt1 | 0.0050 (2) | 0.0040 (2) | 0.0049 (2) | 0 | 0.0009 (1) | 0 |
| Pt2 | 0.0049 (2) | 0.0042 (2) | 0.0049 (2) | 0 | 0.0009 (1) | 0 |
| Dy1 | 0.0020 (2) | 0.0030 (2) | 0.0048 (2) | 0 | 0.0013 (2) | 0 |
| Sb1 | 0.003 (2) | 0.004 (2) | 0.008 (2) | 0 | 0.0003 (10) | 0 |
| Sb1' | 0.005 (2) | 0.005 (2) | 0.008 (2) | 0 | 0.002 (1) | 0 |
| Dy2 | 0.019 (1) | 0.011 (1) | 0.009 (1) | 0 | -0.0003 (9) | 0 |
| Sb2 | 0.021 (9) | 0.019 (9) | 0.016 (8) | 0 | 0.006 (6) | 0 |
| Sb2' | 0.016 (8) | 0.012 (8) | 0.008 (8) | 0 | 0.001 (5) | 0 |
| Dy3 | 0.0057 (2) | 0.0045 (2) | 0.0053 (2) | 0 | 0.0011 (1) | 0 |
| Dy4 | 0.0056 (2) | 0.0043 (2) | 0.0054 (2) | 0 | 0.0010 (1) | 0 |
| Sb3 | 0.0061 (3) | 0.0050 (3) | 0.0056 (3) | 0 | 0.0012 (2) | 0 |
| Sb4 | 0.0057 (3) | 0.0053 (3) | 0.0057 (3) | 0 | 0.0012 (2) | 0 |
| Sb5 | 0.0043 (3) | 0.0048 (3) | 0.0067 (3) | 0 | 0.0013 (2) | 0 |
| Sb6 | 0.0043 (3) | 0.0045 (3) | 0.0064 (3) | 0 | 0.0013 (2) | 0 |

*For an explanation of the anisotropic thermal displacement parameters, see *The International Tables for Crystallography*[39], A. Authier editor, second edition, volume D, pages 231 to 245, John Wiley and Sons, 2014.

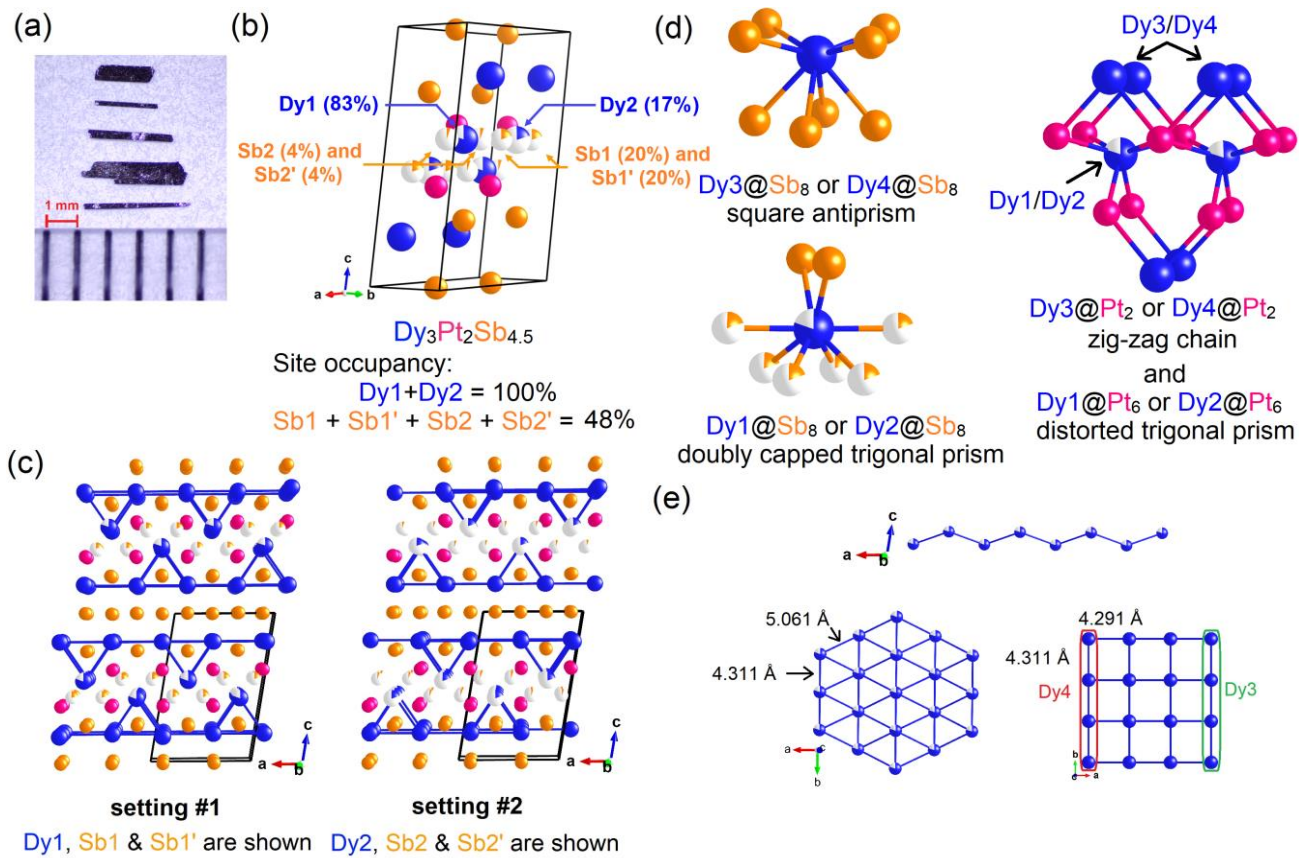

**Figure 1. (a)** Pictures of the $Dy_{3.00(1)}Pt_2Sb_{4.48(2)}$ crystals under an optical microscope. The composition is from single-crystal X-ray diffraction results. **(b)** Unit cell and structure of $Dy_{3.00(1)}Pt_2Sb_{4.48(2)}$ with all disordered sites where blue, red, and orange spheres represent Dy, Pt and Sb atoms, respectively. **(c)** The two different settings of $Dy_{3.00(1)}Pt_2Sb_{4.48(2)}$ with distinct disordered sites of Dy and Sb. **(d)** The coordination environment of Dy represented by setting#1. **(e)** Dy-Dy frameworks represented by setting#1.

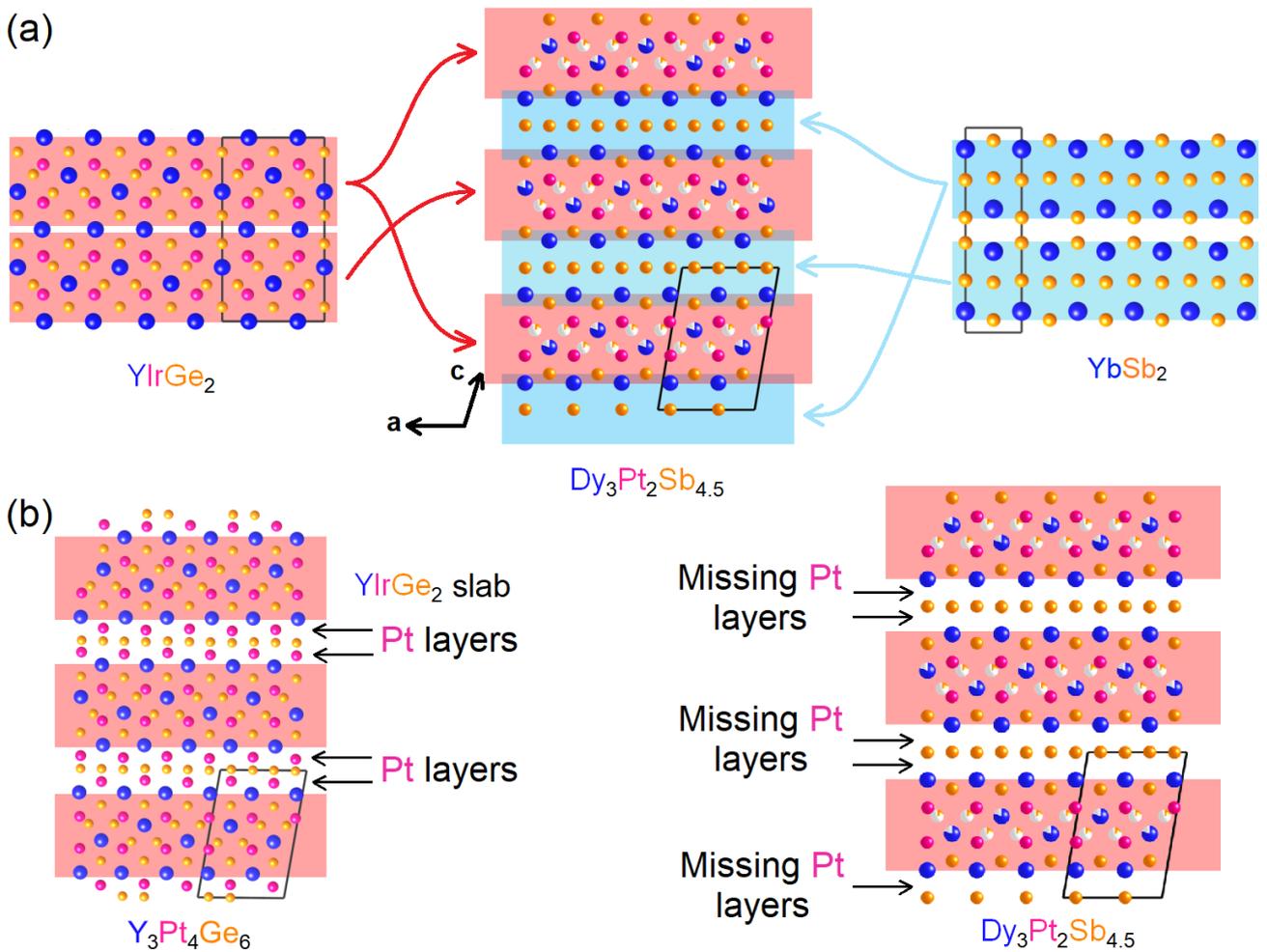

**Figure 2. (a)** The relation among structural motifs of YIrGe$_2$ (left), Dy$_3$Pt$_2$Sb$_{4.48}$ (middle) and YbSb$_2$ (right). The shaded parts with the same color are similar to each other and the arrows indicate the location of the intergrowth. **(b)** Comparison between the structures of Y$_3$Pt$_4$Ge$_6$ and Dy$_3$Pt$_2$Sb$_{4.48}$. The red shades stand for the common YIrGe$_2$ slabs in both structural types. The arrows indicate where the absences of Pt atoms happen.

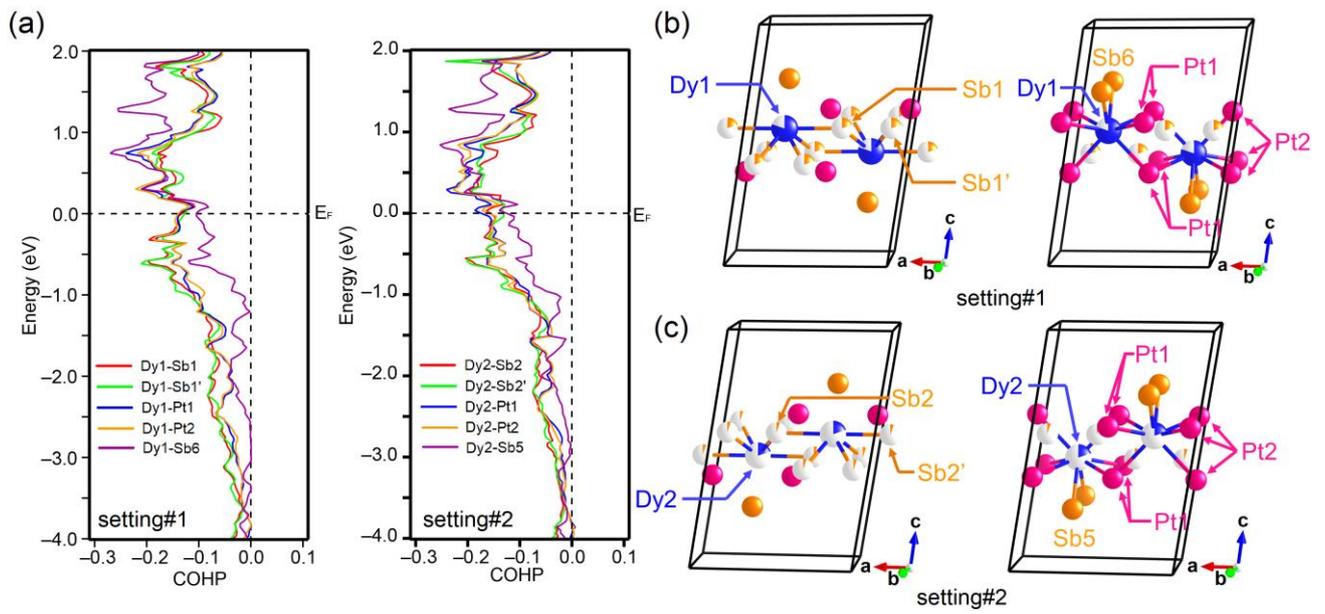

**Figure 3. (a)** The COHP curves for both settings of Dy$_3$Pt$_2$Sb$_{4.48}$. **(b) & (c)** Visualization of related chemical bonds mentioned in **(a)**.

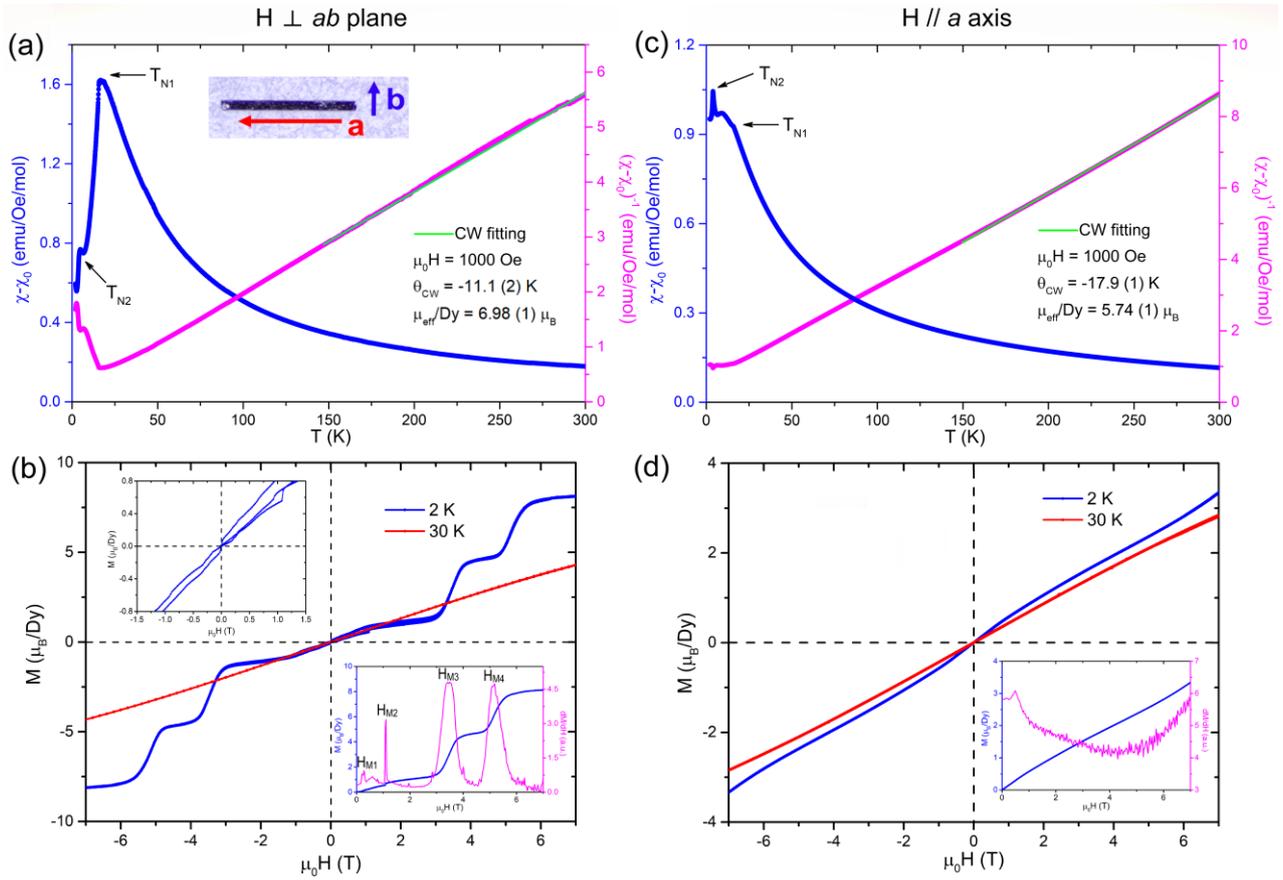

**Figure 4.** The temperature dependence of magnetic susceptibility and inverse magnetic susceptibility when the external magnetic field is **(a)** perpendicular to the ab plane; **(c)** parallel to the *a* axis. The inset of **(a)** is a picture of crystal shown the direction of crystallographic axes. **(Main panel)** The hysteresis loop of $Dy_3Pt_2Sb_{4.48}$ under various temperatures when the external magnetic field is **(b)** perpendicular to the ab plane; **(d)** parallel to the *a* axis. The inset figures in **(b)** are **(top left)** enlarged hysteresis loop at 2 K from -1.5 T to 1.5 T and **(bottom right)** M vs H curve from 0 to 7 T with its first derivative. The inset in **(d)** is the M vs H curve from 0 to 7 T with its first derivative.

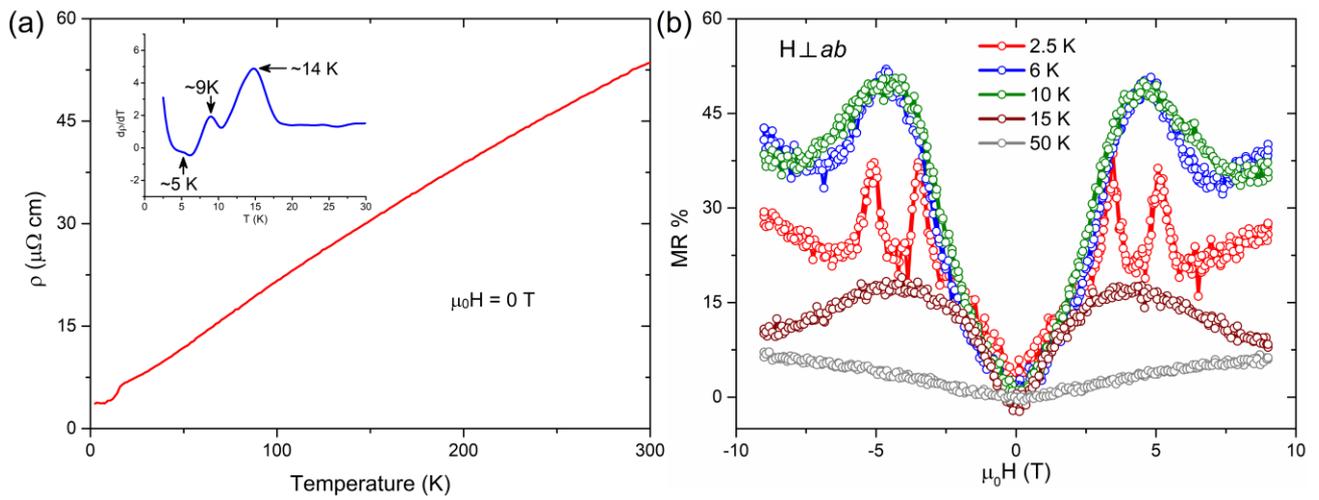

**Figure 5. (a) (Main panel)** Temperature-dependent electrical resistivity with no magnetic field applied. **(Inset)** The first derivative of the main panel figure. **(b)** Magnetoresistance measured from -9 T to 9 T at various temperatures when the magnetic field is applied perpendicularly to the ab plane.

# Supplementary Information

# Crystallographic Disorder and Strong Magnetic Anisotropy in $Dy_3Pt_2Sb_{4.48}$

*Terry Paske,[a] Yingdong Guan,[b] Chaoguo Wang,[a] Curtis Moore,[c] Zhiqiang Mao,[b] Xin Gui [a*]*

[a] Department of Chemistry, University of Pittsburgh, Pittsburgh, PA, 15260, USA
[b] Department of Physics, Pennsylvania State University, University Park, PA, 16801, USA
[c] Department of Chemistry and Biochemistry, The Ohio State University, Columbus, OH, 43210, USA

## Table of Contents

**Table S1.** Comparsison of single crystal structure refinement results for $Dy_{3.00(1)}Pt_2Sb_{4.50(2)}$ at 273 (2) K.

| Refined Formula | $Dy_{3.00(1)}Pt_2Sb_{4.50(2)}$ | No Sb2 and Sb2' | No Sb2/2' and Dy2 |
|---|---|---|---|
| F.W. (g/mol) | 1424.95 | 1424.95 | 1424.95 |
| Space group; Z | $P 2_1/m$; 2 | $P 2_1/m$; 2 | $P 2_1/m$; 2 |
| $a$ (Å) | 8.6252 (8) | 8.6252 (8) | 8.6252 (8) |
| $b$ (Å) | 4.3109 (3) | 4.3109 (3) | 4.3109 (3) |
| $c$ (Å) | 12.968 (1) | 12.968 (1) | 12.968 (1) |
| $\beta$ (º) | 99.609 (3) | 99.609 (3) | 99.609 (3) |
| $V$ (Å³) | 475.43 (7) | 475.43 (7) | 475.43 (7) |
| Extinction Coefficient | 0.00040 (4) | 0.00032 (5) | 0.00000 (13) |
| θ range (º) | 3.090-31.902 | 3.090-31.902 | 3.090-31.902 |
| No. reflections; $R_{int}$ | 23877; 0.0329 | 23877; 0.0329 | 23877; 0.0329 |
| No. independent reflections | 1699 | 1699 | 1699 |
| No. parameters | 91 | 79 | 71 |
| $R_1$: $\omega R_2$ ($I>2\delta(I)$) | 0.0257; 0.0392 | 0.0316; 0.0500 | 0.0845; 0.1405 |
| Goodness of fit | 1.143 | 1.431 | 3.971 |
| Diffraction peak and hole (e⁻/ Å³) | 2.098; -2.257 | 8.992; -2.445 | 58.807; -5.725 |

**Table S2.** The atomic ratios from Energy-Dispersive X-ray Spectroscopy (EDS) result of two different crystals of $Dy_3Pt_2Sb_{4.5}$.

|  | Dy% (Error%) | Pt% (Error%) | Sb% (Error%) |
|---|---|---|---|
| **Number of points** | | Sample1 | |
| 1 | 34.71 (3.95) | 22.71 (7.93) | 42.58 (4.72) |
| 2 | 34.74 (3.88) | 22.15 (7.99) | 43.11 (4.66) |
| 3 | 34.68 (3.83) | 22.02 (7.28) | 43.31 (4.65) |
| 4 | 34.89 (3.91) | 22.30 (7.54) | 42.81 (4.84) |
| 5 | 34.77 (3.92) | 22.45 (8.10) | 42.78 (4.84) |
| **Number of points** | | Sample2 | |
| 1 | 34.81 (3.93) | 22.99 (7.75) | 42.20 (4.90) |
| 2 | 34.86 (3.88) | 22.80 (7.38) | 42.35 (4.88) |
| 3 | 34.75 (3.85) | 22.87 (8.03) | 42.38 (4.87) |
| 4 | 35.06 (3.94) | 22.63 (7.79) | 42.31 (4.86) |
| 5 | 34.99 (3.94) | 22.51 (7.86) | 42.50 (4.84) |
| **Average** | 34.82 (3.90) | 22.54 (7.77) | 42.63 (4.81) |
| **Normalized to Pt** | 3.1 (3) | 2.0 (7) | 3.8 (4) |

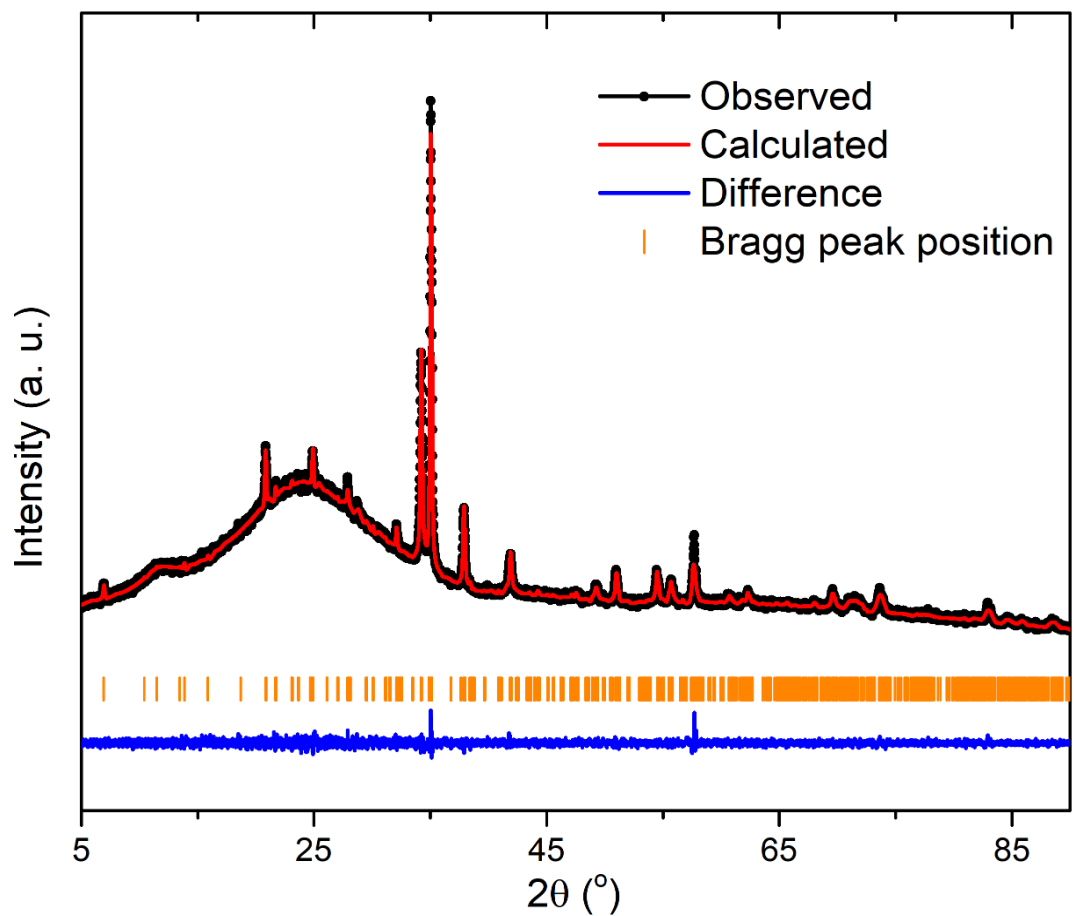

**Figure S1.** Powder XRD pattern of the crushed crystals of $Dy_{3.00(1)}Pt_2Sb_{4.48(2)}$. Blue and red lines stand for calculated and observed patterns, respectively. The crushed crystals show a preferred orientation of (00l) so that the observed peak for, for instance, (005) peak is much higher than calculated. Therefore, a refinement for preferred orientation of (00l) has been conducted.

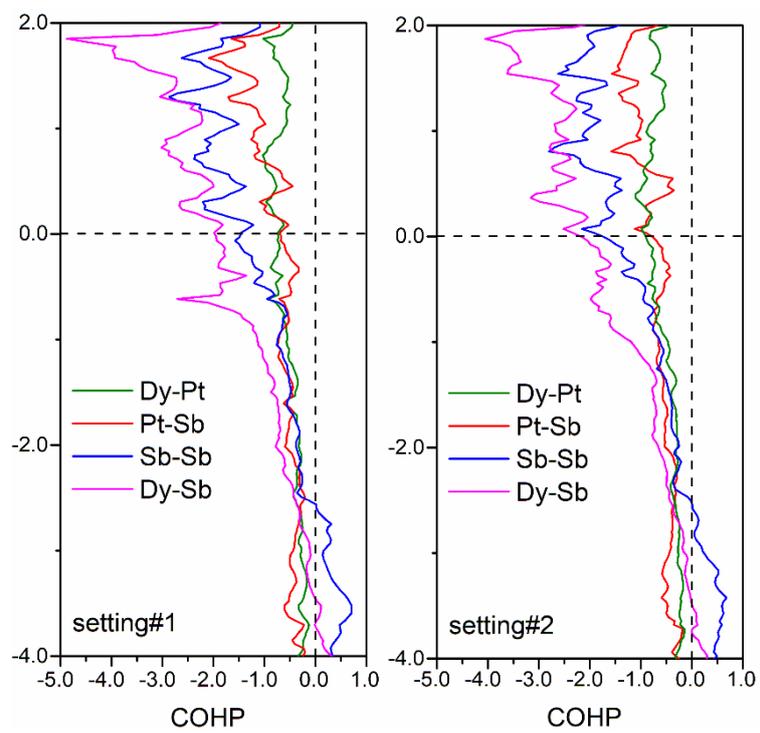

**Figure S2.** COHP curves of $Dy_{3.00(1)}Pt_2Sb_{4.48(2)}$ for both settings.

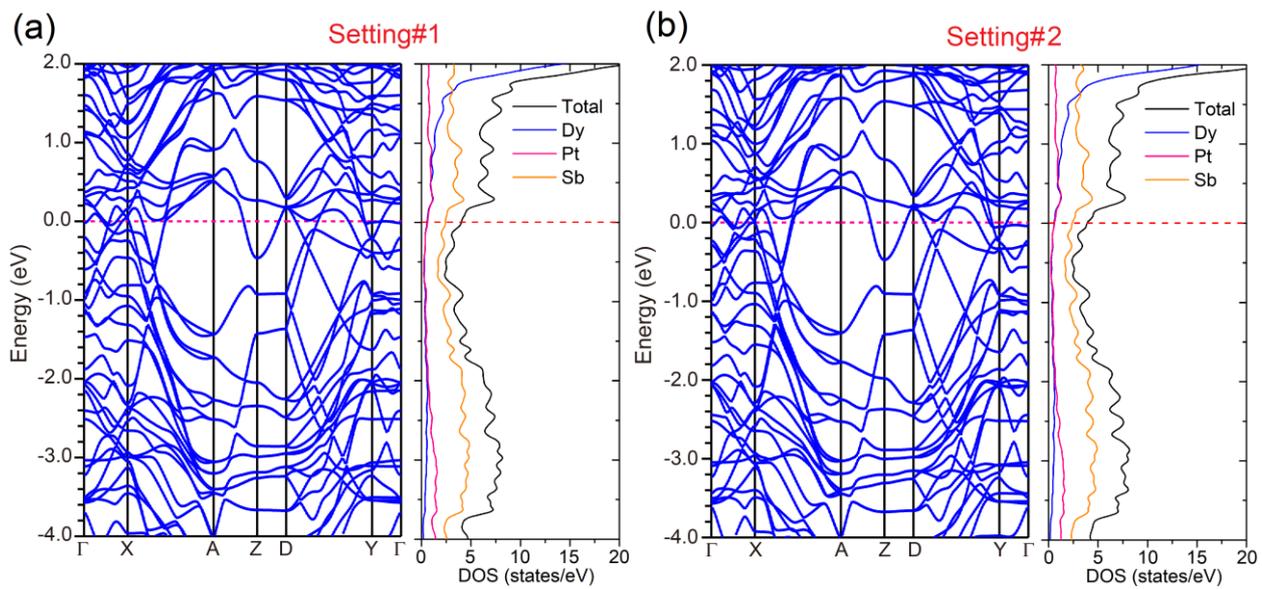

**Figure S3.** Band structure and DOS of $Dy_{3.1(3)}Pt_{2.0(7)}Sb_{3.8(4)}$ for **(a)** setting#1 and **(b)** setting#2.